\documentclass[%
12pt,
 oneocolumn,
 nofootinbib,
 amsmath,amssymb,
 aps,
floatfix,
unsortedaddress
]{revtex4}

\usepackage{graphicx,color}
\usepackage{subfig}
\usepackage{dcolumn}% Align table columns on decimal point
\usepackage{bm}% bold math
\usepackage{graphicx,color,caption}
\usepackage[font=small,justification=Justified,format=plain]{caption}
\usepackage{float}
\usepackage{tikz}
\usetikzlibrary{decorations.markings}
\usetikzlibrary{snakes}
\tikzstyle directed=[postaction={decorate,decoration={markings,
        mark=at position .4 with {\arrow[scale=1]{stealth}}}}]
\usepackage[pdfpagelabels, pdfencoding=auto, psdextra]{hyperref}
\hypersetup{%
 pdfsubject=Paper,
 pdfkeywords={nuclear physics} {Pionless EFT} {Two-Nucleon},
 unicode = true,
 breaklinks = true,
 colorlinks = true,
 linkcolor = blue,
 citecolor = blue,
 menucolor = blue,
 citecolor = blue,
 urlcolor = blue
}

\graphicspath{{/}}

\begin{document}

\title{Radiative Capture $d(\alpha,\gamma)^{6}\mathrm{Li} $ Reaction in Cluster Effective Field Theory}

\author{F. Nazari}
 \email{f.nazari@email.kntu.ac.ir}
 \author{M. Radin}
\email{radin@kntu.ac.ir}
%\thanks{(Corresponding author)}
 \affiliation{Department of Physics, K. N. Toosi University of Technology, P.O.Box 16315-1618, Tehran, Iran
}

\author{M. Moeini Arani }
\email{m.moeini.a@ut.ac.ir}
\affiliation{Malek Ashtar University of Technology, Tehran, Iran}

\begin{abstract}
In this study, we focus on the radiative capture process of the deuteron on alpha particle leading to the formation of $^6{\textrm{Li}}$ in the two-body formalism through the cluster effective field theory(EFT). This was the primitive nuclear reaction producing ${^6 \textrm{Li}}$  a few minutes after the Big Bang. In detail, we outline the calculation of the dominant $E1$ and $E2$ electromagnetic transition amplitudes of $d(\alpha,\gamma ){^6\textrm{Li}}$ up to next-to-leading order. Then we obtain the astrophysical S-factor by fitting it to the experimental data. Finally, we compare the obtained EFT results for the astrophysical S-factor with the other theoretical results.
   
\end{abstract}
\maketitle

%%%%%%%%%%%%%%%%%
\section{\label{Intro}Introduction}
In the standard Big Bang Nucleosynthesis (BBN) framework, the primitive ${^6\textrm{Li}}$ abundance is mainly determined by two nuclear reactions: the
$d(\alpha,\gamma ){^6\textrm{Li}}$ reaction, where deuteron reacts with an alpha particle to produce ${^6\textrm{Li}}$. This reaction leads to the formation of ${^6\textrm{Li}}$ during the primordial nucleosynthesis process \cite {trezzi2017big,serpico2004nuclear}. Conversley, ${^6\textrm{Li}}(p,\alpha ){^3\textrm{He}}$ can destroy ${^6\textrm{Li}}$. In this reaction, ${^6\textrm{Li}}$ reacts with a proton to produce an alpha particle and ${^3\textrm{He}}$. This reaction reduces the abundance of ${^6\textrm{Li}}$ in the early universe. This reaction rate is commonly understood within the BBN energy range and has been extensively researched using various techniques\cite{xu2013nacre, arai2002microscopic, lamia2013updated, fiedler1967wirkungsquerschnitte, spitaleri2006recent}.\\
Consequently, the ${^6\textrm{Li}}$ production reaction has recently been a prime focus of studies,  both experimentally and theoretically \cite{hammache2010high, mukhamedzhanov2011reexamination, tursunov2016theoretical,tursunov2018theoretical, mukhamedzhanov1995astrophysical, cecil1996reaction,langanke1986microscopic, nollett2001six, tursunov2015theoretical, kharbach1998microscopic}. The available experimental
data for this reaction cover the region $100  \textrm{keV}
\leq E_{CM}\leq 4  \textrm{MeV}$ \cite{trezzi2017big,robertson1981observation,anders2014first,mohr1994direct,kiener1991measurements, igamov1999triple}. The low-energy experimental data
have been obtained indirectly from measurements using the
Coulomb breakup process ${^6 \textrm{Li}} +  {^{208} \textrm{Pb}}\rightarrow \alpha +d+  {^{208} \textrm{Pb}} $ \cite{kiener1991measurements}.
The most historical theoretical calculations of the low-energy
cross section for the reaction has been performed within the
framework of the microscopic resonating group method, in the quasi-microscopic potential model and the framework of the multicluster dynamic model \cite {typel1991low, burkova1990possible, ryzhikh1995alpha}.\\
In the present work, we focus on applying cluster effective
field theory (CEFT) formalism as one of the most precise techniques for low-energy nuclear processes for the investigation of gamma capture $d(\alpha,\gamma)^6{\textrm{Li}}$ reaction.
The CEFT is the ideal tool to
analyze the features of halo states with minimal
assumptions \cite{hammer2017effective,ryberg2016cluster}. It describes systems using their effective
degrees of freedom, i.e., core and valence nucleons, and
interactions that are dictated by low-energy constants.
CEFT was formulated in the study of the
shallow p-wave neutron-alpha resonance and applied to other systems, such as the s-wave
alpha-alpha resonance, electromagnetic transitions, and capture reactions. In this paper,
we apply this idea to the $d(\alpha, \gamma){^6\textrm{Li}}$ radiative reaction, following a cluster approach of point-like objects. Our investigation leads to calculating the astrophysical S-factor of the mentioned reaction. \\
The manuscript is organized as follows: In Sec. \ref{Radiative Capture}, the possible and dominant electromagnetic transition of the $d(\alpha, \gamma){^6\textrm{Li}}$ reaction was introduced. Moreover, The effective non-relativistic lagrangian of the system was presented, concluding the scattering and additional terms for the radiative capture process.  The details of the Coulomb interaction were investigated in this section. Sec. \ref{Elastic Scattering } is dedicated to the basic feature of the elastic scattering  $d(\alpha, \alpha)d$ reaction in CEFT. In Sec. \ref{E1 and E2 Transition Amplitudes}, we review the principles of radiative capture reaction. Furthermore, we derive
relevant expressions for the capture
amplitude and cross-section for the $E1$ and $E2$ transitions in this section.
Sec. \ref{Results} collects the numerical results, in particular, the astrophysical S-factor at leading and next-to-leading order.  Our concluding remarks are presented in  Sec. \ref{Conclusion}.

\section{\label{Radiative Capture}Radiative Capture Reaction}

In this section, we first discuss the possible and most prevalent electromagnetic transitions that occur during the radiative capture process of $d(\alpha,\gamma)^{6}\mathrm{Li}$ at low energies. Then, we present the Lagrangian of this reaction based on two-body CEFT.
From a schematic point of view, in a radiative capture reaction, a stationary nucleus in a definite quantum
mechanical state transitions to a lower energy state via the emission of a single photon.
The conservation of angular momentum plays a controlling role in the gamma ray decay process. Both the initial and final states of the nucleus will possess a precise angular momentum and parity, necessitating the photon to connect these two states while preserving both parity and angular momentum. Photons carry a precise amount of angular momentum and possess a specific parity, with the conservation of these properties influencing the characteristics of the emitted photon. The electromagnetic selection rules and multipolarities for nuclear physics are outlined in Table. \ref{tab:1}.
\begin{table}[h]
\begin{center}
\centering
\caption{\label{tab:1} The electromagnetic selection rules. $\Delta l $ and $\Delta \pi$ are angular momentum and parity of the photon.  }
\begin{tabular}{|c|c|c|c|}
\hline
Radiation Type &Name &$\Delta l $&$\Delta \pi$ \\
\hline
\hline
$E1$& Electric dipole &$1 $&yes \\
\hline
$M1$& Magnetic dipole &$1 $&no \\
\hline
$E2$& Electric quadrupole &$2 $&no \\
\hline
$M2$&Magnetic quadrupole &$2 $&yes\\
\hline
\end{tabular}
\end{center}
\end{table}

Generally, the type of photon involved in a transition between nuclei can be determined by considering the properties of photons. First, the parity of the
photon ($\Delta \pi$) is determined by the difference in parities of the two nuclear states. The
photon angular momentum is then constrained within the range of $|l_i-l_f| \leq\Delta l\leq l_i+l_f$, where $l_i$  and $l_f$ are the angular momenta of the initial and final states of the nucleus, respectively.
The multipolarity of the photon is specified by the amount of angular momentum
carried by the photon \cite {siegbahn2012alpha}.

Considering the alpha (spin-zero)  and deuteron (spin-one) particles as point-like particles and 
 taking into account the $l$-wave components
of the two-body $d-\alpha$ system, the possible incoming states for the two-body
$d-\alpha$ system are $\xi=^3\!S_1$, $^3\!P_0$, $^3\!P_1$,
$^3\!P_2$, $^3\!D_1$, $^3\!D_2$, $^3\!D_3$ corresponding to the
relevant spin-angular momenta, $j=0, 1, 2, 3$. 
Considering the alpha (spin-zero)  and deuteron (spin-one) particles as point-like particles and 
 taking into account the $l$-wave components
of the two-body $d-\alpha$ system, the possible incoming states for the two-body
$d-\alpha$ system are $\xi=^3\!S_1$, $^3\!P_0$, $^3\!P_1$,
$^3\!P_2$, $^3\!D_1$, $^3\!D_2$, $^3\!D_3$ corresponding to the
relevant spin-angular momenta, $j=0, 1, 2, 3$. 
Thus, we consider the final ground state of the $^{6}\mathrm{Li}$ nucleus with $J^{\pi}=1^+$ as $^3\!S_1$.
According to the electromagnetic transition rules of nuclear physics,
the $E1$  and $E2$ transitions contribute to the $d(\alpha,\gamma)^{6}\mathrm{Li}$ radiative capture amplitude in the low-energy regime.
Transitions from $P$-waves to $S$-wave, which change parity and result in an angular momentum change of $\Delta l\!=\!1$, are considered $E1$ transitions. 
Transitions from $D$-waves to $S$-wave, which does not change parity but involves an angular momentum change of $\Delta l\!=\!2$, contribute to the $E2$ transition. Therefore, the possible electromagnetic transitions during the radiative capture process in the $d(\alpha,\gamma)^{6}\mathrm{Li}$ reaction at low energies can be outlined as follows
\begin{eqnarray}
^3\!P_0, ^3\!\!P_1, ^3\!\!P_2 &\xrightarrow[]{\,\,E1\,\,}{ ^3\!S_1},\quad\quad\qquad\qquad ^3\!D_1, ^3\!\!D_2, ^3\!\!D_3 &\xrightarrow[]{\,\,E2\,\,}{ ^3\!S_1}.\nonumber
\end{eqnarray}

\subsection{\label{Effective Lagrangian}Effective Lagrangian}

In this study, we consider the deuteron and alpha as point-like particles. Therefore, in the cluster EFT that we construct for the $d-\alpha$ system, the degrees of freedom are only the alpha and deuteron particles, and the $^{6}\mathrm{Li}$ nucleus treats as a bound state of point-like nuclear clusters alpha and deuteron with a binding energy $B=1.47 \,\textrm{MeV}$. Therefore, the effective non-relativistic Lagrangian which describes the dynamics in all possible channels $\xi$, can be written as
\begin{eqnarray}\label{eq:1}
\mathcal{L}^{[\xi]}=\mathcal{L}_{ES}^{[\xi]}+\mathcal{L}_{RC}^{[\xi]},
\end{eqnarray}
where $\mathcal{L}_{ES}^{[\xi]}$ defines the elastic scattering and $\mathcal{L}_{RC}^{[\xi]}$ denotes the additional terms for the radiative capture process.
The Lagrangian $\mathcal{L}_{ES}^{[\xi]}$ is given by \cite{Nazari2023low}
\begin{eqnarray}\label{eq:2}
\mathcal{L}_{ES}^{[\xi]}&\!\!\!=&\!\!\!\phi^{\dagger}(i\partial _0+\frac{\boldsymbol{\nabla}^2}{2m_\alpha})\phi+d_i^{\dagger}(i\partial _0+\frac{\boldsymbol{\nabla} ^2}{2m_d})d_i+\,\eta^{[\xi]}t^{[\xi]^\dagger}\!\Big[i\partial _0\!+\!\frac{\boldsymbol{\nabla} ^2}{2m_t}-\Delta^{[\xi]}\Big]\!t^{[\xi]}\nonumber\\
&&\!\!\!+g^{[\xi]}[t^{[\xi]^\dagger}\!(\phi\, \mathrm{\Pi}^{[\xi]}d)+h.c.] +\,h^{[\xi]}t^{[\xi]^\dagger}\!\Big[(i\partial _0\!+\!\frac{\boldsymbol{\nabla} ^2}{2m_t})^2\Big]t^{[\xi]}
+\cdots,
\end{eqnarray}
where "$\cdots$" stands for the terms with more derivatives and/or
auxiliary fields. The scalar field $\phi$ represents the spinless
$\alpha$ field with a mass of $m_\alpha\! =\! 3727.38 ~\textrm{MeV}$, and the
vector field $d_i\!=\!\varepsilon_i^d d$ indicates the deuteron nucleus
axillary field with a mass of $m_d \!=\!1875.61 ~\textrm{MeV}$. The sign
$\eta^{[\xi]}$ is used to match the sign of the effective range
$r^{[\xi]}$ and reflects the auxiliary character of the dimeron
field.
The coupling constants $\Delta^{[\xi]}$, $g^{[\xi]}$ and $h^{[\xi]}$ for each channel cannot be measured directly, but their renormalized values are determined by matching to the available experimental data of phase shifts as we did in the previous work~\cite{Nazari2023low}.
The dimeron field $t^{[\xi]}$ with a mass of $m_t \!=\! m_d \!+\!
m_{\alpha}$, and projection operator ${\mathrm{\Pi}}^{[\xi]}$ for each channel $\xi$ are
defined as
\begin{eqnarray}\label{eq:3}
t^{[\xi]}=\left\lbrace \begin{array}{lc}
    \bar{t}_i,  &\quad \, \xi={^3\!S_1}\\
    t, & \,\quad \xi={^3\!P_0}\\
    t_{k}, &\,\quad \xi={^3\!P_1} \\
    t_{ij}, & \,\quad \xi={^3\!P_2}\\
    \tilde{t}_j, & \,\quad \xi={^3\!D_1}\\
    \tilde{t}_{kl}, & \,\quad \xi={^3\!D_2}\\
    \tilde{t}_{kji}, &\,\quad \xi={^3\!D_3}
    \!\!\end{array}\right\rbrace,\,\qquad
\mathrm{\Pi}^{{[\xi]}}=\left\lbrace \begin{array}{lc}
    \varepsilon_i^d,  & \,\xi={^3\!S_1}\\
    \sqrt{3}\,\mathcal{P}_i\,\varepsilon_i^d, & \, \xi={^3\!P_0}\\
    \sqrt{3/2}\,\epsilon_{kji}\,\mathcal{P}_j\,\varepsilon_i^d ,&\, \xi={^3\!P_1} \\
    3/\sqrt{5}\,\mathcal{P}_j\,\varepsilon_i^d, & \, \xi={^3\!P_2}\\
    3/\sqrt{2}\,\tau_{ji}\,\varepsilon_i^d, & \, \xi={^3\!D_1}\\
    \sqrt{3/2}\,\epsilon_{ijl}\,\tau_{kj}\,\varepsilon_i^d,  & \, \xi={^3\!D_2}\\
    \sqrt{45/8}\,\tau_{kj}\,\varepsilon_i^d, & \, \xi={^3\!D_3}
    \end{array}\right\rbrace,\,\,\,\qquad
\end{eqnarray}
with the derivative operators, which are introduced as
\begin{eqnarray}\label{eq:4}
\mathcal{P}_i=\frac{\mu}{i}\Big(\frac{\overrightarrow{\nabla_i}}{m_{d}}
 -\frac{\overleftarrow{\nabla_i}}{m_{\alpha}} \Big ),\quad\quad\,
\tau_{ij}
 =\mathcal{P}_i \mathcal{P}_j -\frac{1}{3}\delta_{ij}\mathcal{P}_k\mathcal{P}_k.\quad
\end{eqnarray}

\begin{figure}
\begin{center}
\begin{tikzpicture}
\draw[double,thick](0.0,-2.5)--(2.,-2.5);
\draw[dashed](0.0,-3.9)--(2.,-3.9) node[above]{$$};
\draw[fill=pink] (1.,-3.2) ellipse (6pt and 30pt);
\draw[thick](2.6,-3.2)--(2.8,-3.2);
\draw[thick](2.6,-3.3)--(2.8,-3.3);
\draw[double,thick](3.4,-2.5)--(5.4,-2.5);
\draw[dashed](3.4,-3.9)--(5.4,-3.9) node[above]{$$};
\draw (6.,-3.2) -- (6.2,-3.2);
\draw (6.1,-3.325) -- (6.1,-3.075);
\draw[double,thick](7,-2.5)--(9,-2.5);
\draw[dashed](7,-3.9)--(9,-3.9) node[above]{$$};
\draw [snake=snake] (8,-2.5) -- (8,-3.9);
\draw (9.6,-3.2) -- (9.8,-3.2);
\draw (9.7,-3.325) -- (9.7,-3.075);
\draw[double,thick](10.6,-2.5)--(12.6,-2.5);
\draw[dashed](10.6,-3.9)--(12.6,-3.9) node[above]{$$};
\draw [snake=snake] (11.4,-2.5) -- (11.4,-3.9);
\draw [snake=snake] (11.9,-2.5) -- (11.9,-3.9);
\draw (13.4,-3.2) -- (13.6,-3.2);
\draw (13.5,-3.325) -- (13.5,-3.075);
\draw (13.9,-3.2)circle(.1mm);
\draw (14,-3.2)circle(.1mm);
\draw (14.1,-3.2)circle(.1mm);
\draw[thick](2.6,-6.2)--(2.8,-6.2);
\draw[thick](2.6,-6.3)--(2.8,-6.3);
\draw[double,thick](3.4,-5.5)--(5.4,-5.5);
\draw[dashed](3.4,-6.9)--(5.4,-6.9) node[above]{$$};
\draw (6,-6.2) -- (6.2,-6.2);
\draw (6.1,-6.325) -- (6.1,-6.075);
\draw[double,thick](7,-5.5)--(10,-5.5);
\draw[dashed](7,-6.9)--(10,-6.9) node[above]{$$};
\draw[fill=pink] (9.2,-6.2) ellipse (6pt and 30pt);
\draw [snake=snake] (8.,-5.5) -- (8.,-6.9);
\end{tikzpicture}
\caption{Coulomb ladder diagrams. The single dashed and double lines represent the scalar $\alpha$ and vector deuteron particle, respectively. The wavy lines represent the exchanged photons.}\label{fig:1}
\end{center}
\end{figure}
The one-body currents are considered by coupling the external photon through minimal substitution, $\mathbf{\boldsymbol{\nabla}} \rightarrow \mathbf{\boldsymbol{\nabla}} +
 ieZ\mathbf{A}$ in the Lagrangian $\mathcal{L}_{ES}^{[\xi]}$, where $Z$ is the charge number and $\mathbf{A}$ is the photon field.
The $E1$ two-body currents using the auxiliary fields and corresponding projection operators of Eq.~\ref{eq:3} for incoming $^3\!P_0$, $^3\!P_1$ and $^3\!P_2$ channels can be described as \cite{higa2018radiative}
\begin{eqnarray}\label{eq:5}
\mathcal{L}_{RC}^{[^3\!P_0]}&\!\!\!=&\!\!\!-\sqrt{3}\,\mu\, Q_1\,L_{E1}\,g^{[^3\!P_0]}g^{[^3\!S_1]} \,t\,\bar{t}_i\,\partial_0A_i,
\\
\mathcal{L}_{RC}^{[^3\!P_1]}&\!\!\!=&\!\!\!-\sqrt{3/2}\,\mu\, Q_1\,L_{E1}\,g^{[^3\!P_1]}g^{[^3\!S_1]}\epsilon_{kij} \,t_k\,\bar{t}_i\,\partial_0A_j,
\\
\mathcal{L}_{RC}^{[^3\!P_2]}&\!\!\!=&\!\!\!-3/\sqrt{5}\,\mu\, Q_1\,L_{E1}\,g^{[^3\!P_2]}g^{[^3\!S_1]} \,t_{ij}\,\bar{t}_i\,\partial_0A_j,
\end{eqnarray}
where $\mu$ is the reduced mass of the $d-\alpha$ system. Moreover, the $E2$ two-body currents for incoming $^3\!D_1$, $^3\!D_2$ and $^3\!D_3$ channels are included by the following Lagrangian \cite{braun2019electric}
\begin{eqnarray}\label{eq:8,9,10}
\mathcal{L}_{RC}^{[^3\!D_1]}&\!\!\!=&\!\!\!3/\sqrt{2}\,\mu\, Q_2\,L_{E2}\,g^{[^3\!D_1]}g^{[^3\!S_1]} \,\tilde{t}_j\,\bar{t}_i\Big(\nabla_{\!j}\nabla_{\!i}A_0\!-\!\partial_0(\nabla_{\!j}A_i+\nabla_{\!i}A_j)/2\Big),\\
\mathcal{L}_{RC}^{[^3\!D_2]}&\!\!\!=&\!\!\!\sqrt{3/2}\,\mu\, Q_2\,L_{E2}\,g^{[^3\!D_2]}g^{[^3\!S_1]}\epsilon_{ijl} \,\tilde{t}_{kl}\,\bar{t}_i\Big(\nabla_{\!k}\nabla_{\!j}A_0\!-\!\partial_0(\nabla_{\!k}A_j+\nabla_{\!j}A_k)/2\Big),\\
\mathcal{L}_{RC}^{[^3\!D_3]}&\!\!\!=&\!\!\!\sqrt{45/8}\,\mu \, Q_2\,L_{E2}\,g^{[^3\!D_3]}g^{[^3\!S_1]} \tilde{t}_{kji}\,\bar{t}_i\Big(\nabla_{\!k}\nabla_{\!j}A_0\! -\! \partial_0(\nabla_{\!k}A_{j}\!+\!\nabla_{\!j}A_{k})/2 \Big).
\end{eqnarray}
The effective charges $Q_1=e\mu\big(Z_{d}/m_d\!-\!Z_{\alpha}/m_{\alpha}\big)$,  $Q_2=e\mu^2\big(Z_{\alpha}/m^2_{\alpha}\!+\!Z_{d}/m^2_d\big)$, and $g^{[^3\!S_1]}$, $g^{[\xi]}$ factors are included in the definition of the couplings $L_{E1}$ and $L_{E2}$. Here, $Z_{\alpha}\!=\!2$ and $Z_{d}\!=\!1$ represent the atomic numbers of the alpha particle and deuteron, respectively.

\subsection{\label{Coulomb Interaction} Coulomb interaction}
%\vspace{-0.4cm}
The strength of the Coulomb-photon exchanges in the $d-\alpha$
interaction is quantified by the
dimensionless Sommerfeld parameter $\eta_p=k_C/p=Z_{\alpha}Z_d\,\alpha_{em}\,\mu/p$,
where $k_C$ is the inverse of the Bohr radius of the $d- \alpha$
system, $\alpha_{em}= e^2/4\pi \sim 1/137$ represents the fine
structure constant, and $p$ is the relative momentum of the alpha and deuteron in
the center-of-mass (CM) framework.
Since, each photon-exchange
insertion is proportional to $\eta_p$ so,
in the low-energy scattering region, where the momentum, $p$, is much less than the characteristic momentum scale $k_C$, it is important to take into account the full Coulomb interaction in a non-perturbative manner, as illustrated in Fig. \ref{fig:1}. \\
To consider the contribution of the Coulomb interaction in
the two-body  $d-\alpha$ system, we use the Coulomb Green's function
as follows \cite {goldberger1964collision}. As depicted in Fig.~\ref{fig:1}, the Coulomb Green's
function is linked to the free Green's function through the
integral equation $\hat{G}_C^{\pm}\!=\!\hat{G}_0^{\pm}\!+\!\hat{G}_0^{\pm}\,\hat{V}_C\,\hat{G}_C^{\pm}$,
where $\hat{G}_0^{\pm}\!=\!1/(E\!-\!\hat{H}_0\!\pm \!i\epsilon)$ and $\hat{G}_C^{\pm}\!=\!1/(E\!-\!\hat{H}_0\!-\!\hat{V}_C\!\pm \!i\epsilon)$ are the free and Coulomb Green's functions for the $d-\alpha$
system with $\hat{V}_C\!=\!2\alpha_{em}/r$ and
$\hat{H}_0\!=\!\hat{p}^2/2\mu$ as the repulsive Coulomb
potential between alpha and deuteron and the free-particle
Hamiltonian, respectively. The signs $(\pm)$ correspond to the
retarded and advanced Green's functions. The Coulomb
wave functions and the retarded Green’s
function can be obtained by solving the Schrodinger
equation with the full Hamiltonian $\hat{H}\! = \!\hat{H}_0\! + \!\hat{V}_C$ as~\cite{kong2000coulomb,holstein1999hadronic}
\begin{eqnarray}\label{eq:11,12,13,14}
\chi_p^{(+)}(\textbf{r})&\!\!\!=&\!\!\!\sum _{l=0}^{\infty}(2l+1)P_l(\hat {\textbf{p}}\cdot\hat{\textbf{r}})\chi_p^{(l)}(r),\\\chi_p^{(l)}(r)&\!\!\!=&\!\!\!i^l e^{i\sigma_l}\, \frac{F_l(\eta_p,pr)}{pr},\\
G_C^{(+)}(E,\mathbf{r}',\mathbf{r})&\!\!\!=&\!\!\!\sum _{l=0}^{\infty}(2l+1) P_l(\hat {\textbf{r}}'\cdot\hat{\textbf{r}})G_C^{(l)}(E,r',r),\\
G_C^{(l)}(E,r',r)&\!\!\!=&\!\!\!-\frac{\mu p}{2\pi}\frac{F_l(\eta_p,pr_<)}{pr_<}\frac{H^{(+)}_l(\eta_p,pr_>)}{pr_>},
\end{eqnarray}
where $\sigma_l\!=\!\sqrt{\mathrm{\Gamma}(l+1+i\eta_p)/\mathrm{\Gamma}(l+1-i\eta_p)}$ indicates the pure Coulomb phase shift and $P_l$ denotes the Legendre function. $r_<$ ($r_>$) corresponds to the lesser (greater) of the coordinates $r$, $r'$ and
\begin{eqnarray}\label{eq:15,16}
F_l(\eta_p,\rho)&\!\!\!=&\!\!\!C_l(\eta_p)2^{-l-1}(-i)^{l+1}M_{i\eta_p,l+1/2}(2i\rho),\\
H^{(+)}_l(\eta_p,\rho)&\!\!\!=&\!\!\!(-i)^{l}e^{\pi\eta_p/2}e^{i\sigma_l}W_{-i\eta_p,l+1/2}(-2i\rho),
\end{eqnarray}
with the conventionally defined the Whittaker functions $M_{k,\mu}(z)$
and $W_{k,\mu}(z)$. The normalized constant $C_l(\eta_p)$ is always positive and has the form
\begin{eqnarray}\label{eq:17}
C^2_l(\eta_p)=\frac{2^{2l}C_0^2(\eta_p)\,\prod_{n=1}^l(n^2+\eta_p^2)}{\mathrm{\Gamma}(2l+2)^2}=2^le^{-\pi\eta_p/2}\,\frac{|\Gamma(l+1+i\eta_p)|}{\Gamma(2l+2)},
\end{eqnarray}
where $C_0^2(\eta_p)$, the probability of finding the deuteron and alpha particles at
zero separation, is defined as
\begin{eqnarray}\label{eq:18}
C_0^2(\eta_p)=\chi_{p'}^{(\pm)}(\mathbf{0})\chi_{p}^{*(\pm)}(\mathbf{0})=\frac{2\pi \eta_p}{e^{2\pi \eta_p}-1}.
\end{eqnarray}
The Coulomb Green’s function for $^{6}\mathrm{Li}$ bound state with two-body binding energy $B$ is defined as
\begin{eqnarray}\label{eq:19}
G_C^{(l)}(-B,r',r)=-\frac{i\mu \gamma}{2\pi}\frac{F_l(\eta_{i\gamma},i\gamma r')}{i\gamma r'}\frac{H^{(+)}_l(\eta_{i\gamma},i\gamma r)}{i\gamma r},
\end{eqnarray}
where $\gamma=\sqrt{2\mu B}$ indicates the two-body binding momentum of $^{6}\mathrm{Li}$ and the coordinate space definitions are as $r' < r$.
\begin{figure}
\begin{center}
\begin{tikzpicture}
\draw[fill=red] (-2,1);
\draw[fill=black] (-2,-1);
\draw[fill=red] (0,0)circle(.6mm)  node[below]{};
\draw[fill=red] (3,0)circle(.6mm)  node[below]{};
\draw[fill=black] (5,1);
\draw[fill=black] (5,-1);
\draw [fill=gray](0,-0.15) rectangle +(3,0.3);
\draw[thick,double] (0,0)--(-2,1)node[above]{$d(\frac{p^2}{2m_d},\textbf{p})$};
\draw[dashed](0,0)--(-2,-1) node[below]{$\alpha(\frac{p^2}{2m_\alpha},-\textbf{p})$};
\draw [thick,double](3,0)--(5,1) node[above]{$d(\frac{p'^2}{2m_d},\textbf{p}')$};
\draw [dashed](3,0)--(5,-1) node[below]{$\alpha(\frac{p'^2}{2m_\alpha},-\textbf{p}')$};
\draw[fill=pink] (-1.25,0) ellipse (8pt and 30pt);
%\draw[fill=pink] (-1.3,-1) rectangle +(0.4,2);
%\draw[fill=gray] (1.5,0) circle (0.4cm);
\draw[ball color=pink] (1.5,0) circle (0.4cm);
\draw[fill=pink] (4.25,0) ellipse (8pt and 30pt);
\node[text width=3cm] at (1.8,-2)
    {$-iT_{CS}^{[\xi]}(E,\textbf{p},\textbf{p}')$};
\end{tikzpicture}
\end{center}
\begin{center}
\begin{tikzpicture}
\draw [fill=gray](0,-0.15) rectangle +(3,0.3);
\draw[ball color=pink] (1.5,0) circle (0.4cm);
\draw[thick] (3.9,0)--(4.1,0);
\draw[thick] (3.9,-0.05)--(4.1,-0.05);
\draw [fill=gray](4.70,-0.15) rectangle +(3,0.3);
\draw[dashed](9,0)--(9.8,0) node[above]{$$};
\draw[dashed](11,0)--(11.8,0) node[above]{$$};
\draw[double,thick] (11,0) -- (11,0) arc (0:180:.6cm);
\draw[dashed](9.8,0) -- (9.8,0) arc (180:360:.6cm);
\draw [fill=gray](8.8,-0.15) rectangle +(1,0.3);
\draw [fill=gray](11,-0.15) rectangle +(3,0.3);
\draw[ball color=pink] (12.5,0) circle (0.4cm);
\draw[fill=pink] (10.4,0) ellipse (6pt and 22pt);
\draw (8.1,0) -- (8.3,0);
\draw  (8.2,-0.125) -- (8.2,0.125);
\end{tikzpicture}
\caption {The amplitude of the $d-\alpha$ elastic scattering.
The thick line is the bare dimeron propagator and the thick dashed line with a filled circle represents the full dimeron
propagator. All remained notations are the same as Fig.~\ref{fig:1}.}\label{fig:2}
\end{center}
\end{figure}
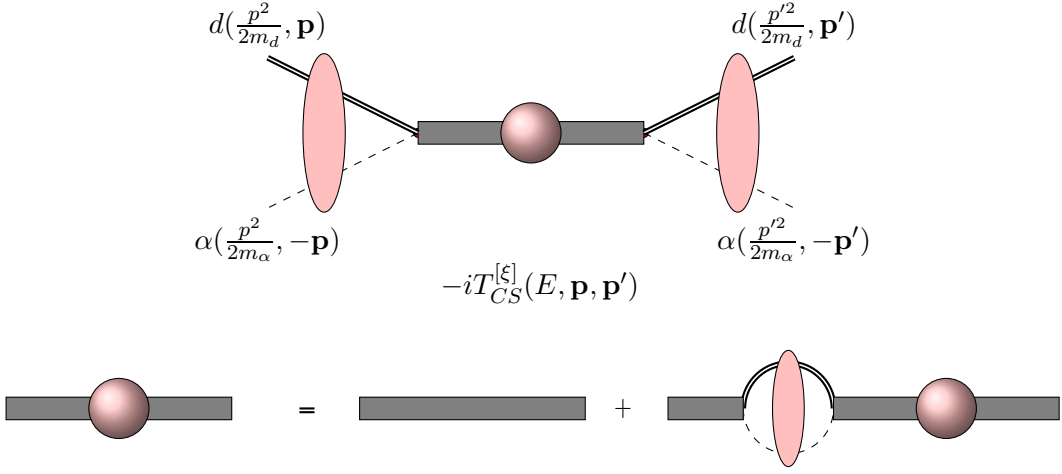

\section{\label{Elastic Scattering }Elastic Scattering }
The elastic scattering amplitude for
two particles  interacting via short-range strong and long-range
Coulomb interactions in the CM framework can be written as~\cite{higa2008alphaalpha}
\begin{equation}\label{eq:20}
T(\textbf{p}',\textbf{p};E)=T_C(\textbf{p}',\textbf{p};E)+T_{CS}(\textbf{p}',\textbf{p};E),
\end{equation}
where $T_{C}$ indicates the pure Coulomb scattering amplitude and
$T_{CS}$ represents the scattering amplitude for the strong
interaction in the presence of the Coulomb interaction with
$E=p^2/2\mu$ as the CM energy of the system. $\textbf{p}$
and $\textbf{p}'$ denote the relative momentum of incoming and
outgoing particles, respectively.
These amplitudes can be expressed in the partial wave decomposition as~\cite{kong2000coulomb}
\begin{eqnarray}\label{eq:21}
T_C(\textbf{p}',\textbf{p};E)=\sum_{l=0}^{\infty}
(2l+1) T_C^{[l]}
P_l(\hat{\mathbf{p}}'\cdot\hat{\mathbf{p}})
=-\frac{2
\pi}{\mu} \sum_{l=0}^{\infty} (2l+1)
\frac{e^{2i\sigma_l}-1}{2ip}P_l(\hat{\mathbf{p}}'\cdot\hat{\mathbf{p}}),
\end{eqnarray}
and
\begin{equation}\label{eq:22}
T_{CS}(\mathbf{p}',\mathbf{p};E)=\sum_{l=0}^{\infty}(2l+1)T^{[l]}_{CS}(p)\,e^{2i \sigma_l}P_l(\hat{\mathbf{p}}'\cdot \hat{\mathbf{p}}),
\end{equation}
with
\begin{eqnarray}\label{eq:23}
T^{[l]}_{CS}(p)=-\frac{2\pi}{\mu}\frac{1}{p (\textrm{cot}\delta_l-i)},\quad\quad\,\,\,
\end{eqnarray}
where $\delta_l$ denotes the Coulomb-corrected phase shift.

At low-energy $d-\alpha$ elastic scattering, the $S$-, $P$- and $D$-wave channels
($\xi={^3\!S_1}$, ${^3\!P_0}, {^3\!P_1}, {^3\!P_2}, {^3\!D_1}, {^3\!D_2},$ ${^3\!D_3}$) dominantly contribute in the elastic scattering cross section.
The CEFT diagram of the $d-\alpha$ elastic scattering
amplitude is shown in Fig. \ref{fig:2}. According to this diagram, the on-shell Coulomb-subtracted EFT amplitudes for each channel $\xi$ can be obtained as~\cite {Nazari2023low}
\begin{eqnarray}\label{eq:24}
-i(2l+1)T_{CS}^{[\xi]}(p)= -ig^{[\xi]^2}D^{[\xi]}(E,\textbf{0}) C^2_0(\eta_p)W_l(\eta_p),
\end{eqnarray}
with the full dimer
propagator for channel $\xi$ as
\begin{equation}\label{eq:25}
D^{[\xi]}(E,\textbf{0})=\frac{\eta^{[\xi]}}{E\!-\!\mathrm{\Delta}^{[\xi]}\!+\!h^{[\xi]}E^2-\!\frac{1}{2l+1}\eta^{[\xi]}g^{[\xi]^2}J_l(E)}.
\end{equation}
On the other hand, the on-shell Coulomb-subtracted amplitude $T^{[\xi]}_{CS}$ can usually be expressed
in terms of a modified effective range expansion (ERE) as \cite{ando2016elastic}
\begin{eqnarray}\label{eq:26}
T^{[\xi]}_{CS}(p) &\!\!\!=&\!\!\!-\frac{2\pi}{\mu}\frac{C_0^2(\eta_p)\,W_l(\eta_p)}{K^{[\xi]}(p)-H_l(\eta_p)},\quad
\end{eqnarray}
with
\begin{eqnarray}\label{eq:27,28,29}
W_l(\eta_p) &\!\!\!=&\!\!\!\frac{k_C^{2l}}{(l!)^2}\,\prod^l_{n=0}(1+\frac{n^2}{\eta_p^2}),
\\
H_l(\eta_p)&\!\!\!=&\!\!\!2k_C W_l(\eta_p)H(\eta_p),\\
H(\eta_p)&\!\!\!=&\!\!\!\psi(i\eta_p)+\frac{1}{2i\eta_p}-\ln(i\eta_p),
\end{eqnarray}
where the function $\psi$ is the logarithmic derivative of
 Gamma function. The function $K^{[\xi]}(p)$ represents the short-range strong interaction which is obtained in terms
of the effective range parameters as \cite{bethe1949theory}
\begin{eqnarray}\label{eq:30}
K^{[\xi]}(p)=-\frac{1}{a^{[\xi]}}+\frac{1}{2}r^{[\xi]}\,p^2+\frac{1}{4}s^{[\xi]}\,p^4+\cdots,
\end{eqnarray}
with $a^{[\xi]}$, $r^{[\xi]}$ and $s^{[\xi]}$ as the scattering length, effective range, and shape parameter, respectively.

The fully dressed bubble $J_{l}$ in Eq. ~\ref{eq:25}, is divergent and should be regularized. We can regularize the divergence by dividing the integral $J_l$ into two finite and infinite parts as $J_l=J_l^{fin}
 + J_l^{div}$. The finite part is obtained as~\cite {ando2007low, kaplan1998two}
\begin{eqnarray}\label{eq:31}
J^{fin}_l(p)=-\frac{\mu}{2\pi}H_l(\eta_p).
\end{eqnarray}

The divergent parts absorb the  $\mathrm{\Delta}^{[\xi]}$, $g^{[\xi]}$ and $h^{[\xi]}$
parameters by introducing the renormalized parameters $\mathrm{\Delta}_{\!R}^{[\xi]}$, $g_{\!R}^{[\xi]}$ and $h_{\!R}^{[\xi]}$.
The details of these regularization and renormalization for all channels are presented in our previous paper \cite{Nazari2023low}. Thus, comparing Eq. \ref{eq:24} to Eq.  \ref{eq:26} yields
\begin{eqnarray}\label{eq:32}\mathrm{\Delta}_{\!R}^{[\xi]}=-\frac{\mu\eta^{[\xi]}g_{\!R}^{[\xi]^2}}{(2l+1)2\pi a^{[\xi]}},\qquad
g_{\!R}^{[\xi]^2}=-\frac{(2l+1)2\pi}{\mu^2\eta^{[\xi]}r^{[\xi]}},\qquad
h_{\!R}^{[\xi]}=-\frac{\mu^3g_{\!R}^{[\xi]^2} s^{[\xi]}}{(2l+1)2\pi}.
\end{eqnarray}
At the binding energy of the ground state of $^{6}\mathrm{Li}$, the amplitude should have a pole at the binding momentum $i\gamma$. Thus, we have
\begin{equation}\label{eq:33}
-\frac{1}{a^{[^3S_1]}}-\frac{1}{2}r^{[^3S_1]}\gamma^2\!+\frac{1}{4}s^{[^3S_1]}\gamma^4\!+...\!-\! H_0(i\gamma)=0.
\end{equation}
By imposing this condition, the effective range parameter $a^{[^3S_1]}$ is related to other effective range parameters which can be fixed by available experimental phase shift data of the elastic $d-\alpha$ scattering.
The renormalization constant of the $^{6}\mathrm{Li}$ wave function which is treated as a bound state of alpha and deuteron clusters is defined by the dressed $S$-wave dimer propagator as
\begin{eqnarray}\label{eq:34}
\frac{1}{\mathcal{Z}}&\!\!\!=&\!\!\!\frac{\partial D^{[^3S_1]}(E,\mathbf{0})^{-1}}{\partial E}\Bigg|_{E=-B}\nonumber\\&\!\!\!=&\!\!\!-\frac{g^{[^3S_1]^2 }\mu^2}{2\pi p} \frac{\partial }{\partial p}\Big(\frac{1}{2}r^{[^3S_1]}(p^2+\gamma^2)\!+\frac{1}{4}s^{[^3S_1]}(p^4-\gamma^4)+...\!-\! H_0(\eta_p)+H_0(i\gamma)\Big)\Big|_{p=i\gamma}.\,
\end{eqnarray}

\section{\label{E1 and E2 Transition Amplitudes}$E1$ and $E2$ Transition Amplitudes}

In this section, we focus on the calculation of the $E1$ and $E2$ transition amplitudes for the capture process $d(\alpha,\gamma) ^6{\textrm{Li}}$.  First, the Feynman diagrams of one- and two-body currents that contribute to the dominant transitions ($E1$ and $E2$) are presented. Then, based on the Feynman rules, the transition amplitudes of diagrams for all possible partial waves $\xi$, are presented. As has been mentioned, $\alpha$ and $d$ are considered as the point-like nuclei and the ${^6\mathrm{Li}}$ is the two-body cluster bound state constructed by $\alpha$ and $d$.
We assign the initial CM momentum $\mathbf{p}$ for the $d-\alpha$ system and the outgoing photon in the final state denoted by $\mathbf{k}$.

In the low-energy regime, $p\! \leq \! k_C \sim 18~ \textrm{MeV}$, the
on-shell CM momentum of the system is scaled as low-momentum $Q$.
The high-momentum scale is set by the lowest energy degrees of
freedom that has been integrated out. According to the fact that
there is no existing explicit pions and any deuteron deformation,
the high-momentum scale $\mathrm{\Lambda}$ has been chosen between the pion
mass, $m_\pi\sim 140$ MeV and the momentum corresponding to the
deuteron binding energy, $B_d$ i.e., $\sqrt{2m_d B_d} \sim 90$ MeV.
Around the $p\sim k_C\sim18$ MeV, the expansion parameter of the
current EFT is estimated to be of the order $1/5$. By increasing the energy,
the expansion deteriorates and the precision of our EFT prediction will
be questionable for $E_{CM}=p^2/(2\mu) >3.3$ MeV. The
Sommerfeld parameter $\eta_p$ is enhanced by decreasing the energy.
So, $\eta_p$ would be large around $p\!\lesssim\! k_C$ and the elastic
scattering and capture amplitudes require nonperturbative treatment of the
Coulomb photons.

\begin{figure}[!h]
\begin{center}
\begin{tikzpicture}

\draw[double,thick](8,0)--(10.5,-1);
\draw[dashed](8,-2)--(10.5,-1) ;
\draw[fill=pink] (8.6,-1)  ellipse (6pt and 28pt);
\node[text width=3cm] at (11.8,-2.5)
 {$(a_1)$};
\draw[fill=pink] (9.7,-1) ellipse (6pt and 20pt);
\draw[fill=white] (10.7,-1) circle (0.3cm);
\draw(10.5,-1.2)--(10.9,-0.8) ;
\draw(10.5,-0.8)--(10.9,-1.2) ;
\draw [snake=snake] (9.2,-0.4) -- (9.7,0.4);
%-----------------------------------------------
\draw[double,thick](8,-3)--(10.5,-4);
\draw[dashed](8,-5)--(10.5,-4) ;
\draw[fill=pink] (8.6,-4)  ellipse (6pt and 28pt);
\node[text width=3cm] at (11.8,-5.5)
 {$(a_2)$};
\draw[fill=pink] (9.7,-4) ellipse (6pt and 20pt);
\draw[fill=white] (10.7,-4) circle (0.3cm);
\draw(10.5,-4.2)--(10.9,-3.8) ;
\draw(10.5,-3.8)--(10.9,-4.2) ;
\draw [snake=snake] (9.2,-4.5) -- (9.7,-5.3);
%------------------------------------------------
\draw[double,thick](0,0)--(1.2,-1);
\draw[dashed](0,-2)--(1.2,-1) ;
\draw[fill=pink] (0.55,-1) ellipse (6pt and 28pt);
\node[text width=3cm] at (4.2,-2.5)
 {$(b_1)$};
\draw [fill=gray](1.2,-0.85) rectangle +(1.6,-0.3);
\draw[ball color=pink] (2,-1) circle (0.4cm);
\draw[double,thick] (2.8,-1) to [bend left=80] (5.5,-1);
\draw[dashed] (5.5,-1) to [bend left=80] (2.8,-1);
\draw[fill=pink] (3.6,-1) ellipse (6pt and 28pt);
\draw[fill=pink] (4.8,-1)ellipse (6pt and 28pt);
\draw[fill=white] (5.7,-1) circle (0.3cm);
\draw(5.5,-0.8)--(5.9,-1.2) ;
\draw(5.5,-1.2)--(5.9,-0.8) ;
\draw [snake=snake] (4.5,0.5) -- (4,-0.2);
%------------------------------------------------------
\draw[double,thick](0,-3)--(1.2,-4);
\draw[dashed](0,-5)--(1.2,-4) ;
\draw[fill=pink] (0.55,-4) ellipse (6pt and 28pt);
\node[text width=3cm] at (4.2,-5.5)
 {$(b_2)$};
\draw [fill=gray](1.2,-3.85) rectangle +(1.6,-0.3);
\draw[ball color=pink] (2,-4) circle (0.4cm);
\draw[double,thick] (2.8,-4) to [bend left=80] (5.5,-4);
\draw[dashed] (5.5,-4) to [bend left=80] (2.8,-4);
\draw[fill=pink] (3.6,-4) ellipse (6pt and 28pt);
\draw[fill=pink] (4.8,-4)ellipse (6pt and 28pt);
\draw[fill=white] (5.7,-4) circle (0.3cm);
\draw(5.5,-3.8)--(5.9,-4.2) ;
\draw(5.5,-4.2)--(5.9,-3.8) ;
\draw [snake=snake] (4,-4.8) -- (4.5,-5.5);

\draw[double,thick](0,-6)--(1.2,-7);
\draw[dashed](0,-8)--(1.2,-7) ;
\draw[fill=pink] (0.55,-7) ellipse (6pt and 28pt);
\node[text width=3cm] at (4.2,-8.5)
 {$(b_3)$};
\draw [fill=gray](1.2,-6.85) rectangle +(1.6,-0.3);
\draw[ball color=pink] (2,-7) circle (0.4cm);
\draw[double,thick] (2.8,-7) to [bend left=80] (5.5,-7);
\draw[dashed] (5.5,-7) to [bend left=80] (2.8,-7);
\draw[fill=pink] (3.6,-7) ellipse (6pt and 28pt);
\draw[fill=pink] (4.8,-7)ellipse (6pt and 28pt);
\draw[fill=white] (5.7,-7) circle (0.3cm);
\draw(5.5,-6.8)--(5.9,-7.2) ;
\draw(5.5,-7.2)--(5.9,-6.8) ;
\draw [snake=snake] (2.8,-6.8) -- (3.2,-6.1);

\draw[double,thick](8,-6)--(9.2,-7);
\draw[dashed](8,-8)--(9.2,-7) ;
\draw[fill=pink] (8.55,-7) ellipse (6pt and 28pt);
\node[text width=3cm] at (11.8,-8.5)
 {$(c)$};
\draw [fill=gray](9.2,-7.15) rectangle +(1.8,0.3);
\draw[ball color=pink] (10.1,-7) circle (0.4cm);
\draw [fill=black](11,-7.25) rectangle +(0.5,0.5);
\draw[fill=white] (11.8,-7) circle (0.3cm);
\draw(11.6,-6.8)--(12.0,-7.2) ;
\draw(11.6,-7.2)--(12.0,-6.8) ;
\draw [snake=snake] (11.2,-7) -- (11.6,-6.1);
\end{tikzpicture}
\caption{Diagrams for the radiative capture process $d(\alpha,\gamma)^6{\textrm{Li}}$. The dashed and double lines represent the scalar $\alpha$ and vector deuteron particles respectively. The wavy line indicates the outgoing photon. The square depicts the two-body currents and the $ \bigotimes$  is the final bound state symbol.}\label{fig:3}
\end{center}
\end{figure}
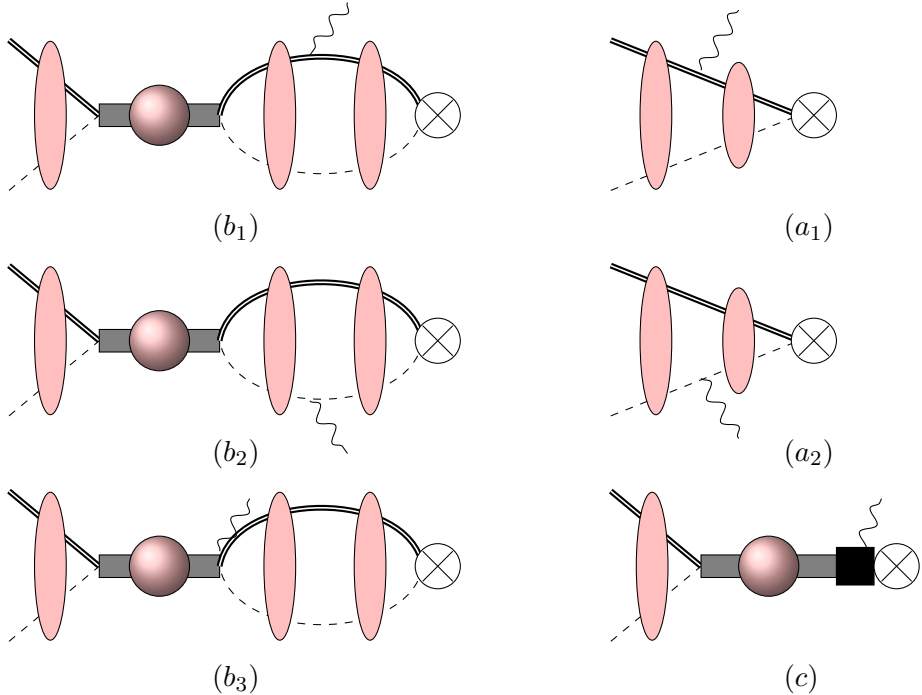
According to the power counting, $\mu\sim \Lambda^3/Q^2$ and $\gamma\sim3Q$. Thus, the energy-momentum conservation and EFT power counting yield
\begin{equation}\label{eq:35}
k=\frac{p^2+\gamma^2}{2\mu}\sim \frac{Q^3}{\Lambda^2}\ll Q\sim p.
\end{equation}
Thus, in the loop calculation, we can use this approximation:
\begin{equation}\label{eq:36}
E_d+k_0+q_0-\frac{(\mathbf{q}+\mathbf{k}+\mathbf{p})^2}{2m_d}\approx E_d+k_0+q_0-\frac{(\mathbf{q}+\mathbf{p})^2}{2m_d}\sim\frac{Q^2}{\mu}\sim\frac{Q^4}{\Lambda^3},
\end{equation}
with $E_d=p^2/(2m_d)$. $(q_0,\mathbf{q})$ and $(k_0,\mathbf{k})$ are the loop and photon energy-momentum, respectively. This approximation corresponds to the zero-recoil of the final bound state $^{6}\mathrm{Li}$.
Consequently, we disregard the recoil effect in our calculations up to next-to-next-to-leading order ($\mathrm{N}^{2}\mathrm{LO}$).

The Feynman diagrams that contribute to the $d(\alpha,\gamma)^6{\textrm{Li}}$ process are shown in Fig. \ref{fig:3}. The diagrams $a_1$ and $a_2$ are the possible transitions including only Coulomb interactions for the incoming charged particles. The diagrams $b_1$, $b_2$, and $b_3$ involve the initial state short-range interactions. In these diagrams, the external photon minimally couples to one of the single-charged particle lines. The last diagram, $c$, denotes the contribution of the two-body current in the gamma capture process.
The $E1$ transition amplitude through all possible incoming $P-$ waves is written as
\begin{eqnarray}\label{eq:37}
M_{E1}(\mathbf{p})=\!\!\!\!\!\!\!\!\!\sum_{\xi={^3\!P_0}, {^3\!P_1}, {^3\!P_2}}\!\!\!\!\!\!\!\Big(M_{E1,a_{1,2}}^{[\xi]}(\mathbf{p})\!+M_{E1,b_{1,2}}^{[\xi]}(\mathbf{p})\!+M_{E1,b_3}^{[\xi]}(\mathbf{p})\!+M_{E1,c}^{[\xi]}(\mathbf{p})\Big).
\end{eqnarray}
As shown in the following, this amplitude leads to
\begin{eqnarray}\label{eq:38}
M_{E1}(\mathbf{p})=\mathcal{M}_{E1}(p)\,\varepsilon^d_i\,\varepsilon^{Li*}_{j} (\boldsymbol{\varepsilon}^{\gamma*}\cdot\hat{\mathbf{p}}),
\end{eqnarray}
where $\boldsymbol{\varepsilon}^{\gamma}$ denotes the polarization vector of the outgoing photon and $\boldsymbol{\varepsilon}^d$ and $\boldsymbol{\varepsilon}^{Li}$ are spin polarization vectors of $d$ and $^{6}\mathrm{Li}$, respectively. The $E1$ transition amplitudes from the initial ${^3\!P_0},{ ^3\!P_1},$ and $ {^3\!P_2}$ states to the final bound state $^3\!S_1$, corresponding to the diagrams $a_1$, $a_2$, $b_1$, $b_2$, $b_3$ and $c$ are obtained as
\begin{eqnarray}\label{eq:39}
M_{E1,a_{1,2}}^{[\xi]}(\mathbf{p})&\!\!\!\!=&\!\!\!\!\frac{i}{\mu}Q_{1}\,g^{[^3S_1]}\sqrt{\mathcal{Z}}  \,\varepsilon^d_i\,\varepsilon^{Li*}_{j}\,\varepsilon^{\gamma*}_k  \!\!\int\! d^3 r\, G_C^{(0)}(-B,0,r)\nabla_{\!k}\!\big[3P_1(\mathbf{\hat{p}}\cdot\mathbf{\hat{r}})\chi^{(1)}_p(r)\big]%\nonumber\\&\!\!\!=&\!\!\!\frac{3}{2\pi }Q_{\textrm{eff}}\,g^{[^3S_1]}\sqrt{\mathcal{Z}}\,   \,\varepsilon^d_i\,\varepsilon^{Li*}_{j}\, \varepsilon^{\gamma*}_k\,\Gamma (1\!+\!k_C/\gamma) e^{ i\sigma_1(\eta_{p})} \,\hat{\mathbf{p}}_l  \nonumber\\
%&&\!\!\!\times\!\int \! d^3 r\,\frac{ W_{-k_C/\gamma,1/2}(2\gamma r)}{r^2}\Big(r\delta_{lk}+r_lr_k\frac{\partial}{\partial r}\Big)\frac{F_1(\eta_p ,pr)}{pr^2}
\nonumber\\&\!\!\!\!=&\!\!\!\!2\,Q_{1}\,g^{[^3S_1]}\sqrt{\mathcal{Z}}\,C_0(\eta_p)\,e^{ i\sigma_1}  \mathcal{A}1(p)   \,\varepsilon^d_i\,\varepsilon^{Li*}_{j}\,(\boldsymbol{\varepsilon}^{\gamma*}\cdot\hat{\mathbf{p}}),
\end{eqnarray}
\begin{eqnarray}\label{eq:40}
M_{E1,b_{1,2}}^{[\xi]}(\mathbf{p})&\!\!\!\!=&\!\!\! \frac{1}{\mu}Q_{1}\,g^{[^3S_1]}\sqrt{\mathcal{Z}}\,   \,\varepsilon^d_i\,\varepsilon^{Li*}_{j}\,\varepsilon^{\gamma*}_k \frac{3T^{[\xi]}_{CS}(p)e^{i\sigma_1}}{C_0(\eta_p)W_1^{1/2}(\eta_p)}\hat{\mathbf{p}}_l \nonumber\\
&&\!\!\!\times\!\int\! d^3 r\,G_C^{(0)}(-B,0,r) \lim_{\!\mathbf{r}''\rightarrow 0} \! \mathbf{\nabla}_{\!k}\mathbf{\nabla}^{''}_{\!l}[3P_1(\hat{\mathbf{r}}\cdot \hat{\mathbf{r}}'')G_C^{(1)}(E,r,r'')] \nonumber\\
&\!\!\!\!=&\!\!\!2\, Q_{1}\,g^{[^3S_1]}\sqrt{\mathcal{Z}}\, \frac{C_0(\eta_p)W_1^{1/2}(\eta_p)}{K^{[\xi]}(p)-H_1(\eta_p)}\,e^{ i\sigma_1} \mathcal{B}1(p)\,\varepsilon^d_i\,\varepsilon^{Li*}_{j}\,(\boldsymbol{\varepsilon}^{\gamma*}\cdot\hat{\mathbf{p}}),
\\\nonumber\\
M_{E1,b_3}^{[\xi]}(\mathbf{p})&\!\!\!\!\!=&\!\!\!\! iQ_{1}\,g^{[^3S_1]}\sqrt{\mathcal{Z}}\,   \,\varepsilon^d_i\,\varepsilon^{Li*}_{j}\,\varepsilon^{\gamma*}_k \,G_C^{(0)}(-B,0,0) \frac{3T^{[\xi]}_{CS}(p)e^{i\sigma_1}}{C_0(\eta_p)W_1^{1/2}(\eta_p)}\hat{\mathbf{p}}_k\nonumber\\&\!\!\!\!\!=&\!\!\!\!-\frac{6\pi i }{\mu}Q_{1}\,g^{[^3S_1]}\sqrt{\mathcal{Z}}\,\frac{C_0(\eta_p)W_1^{1/2}(\eta_p)}{K^{[\xi]}(p)-H_1(\eta_p)}\,e^{ i\sigma_1} J_0  (i\gamma)\,\varepsilon^d_i\,\varepsilon^{Li*}_{j}\,(\boldsymbol{\varepsilon}^{\gamma*}\cdot\hat{\mathbf{p}}),\\\nonumber\\
M_{E1,c}^{[\xi]}(\mathbf{p})&\!\!\!=&\!\!\!i\mu k_0Q_{1}\,g^{[^3S_1]}\sqrt{\mathcal{Z}}\,\varepsilon^d_i\,\varepsilon^{Li*}_{j}\,\varepsilon^{\gamma*}_k  L_{E1}\, \frac{3T^{[\xi]}_{CS}(p)e^{i\sigma_1}}{C_0(\eta_p)W_1^{1/2}(\eta_p)}\,\hat{\mathbf{p}}_k\nonumber\\&\!\!\!=&\!\!\!
-6i\pi  k_0L_{E1}\,Q_{1}\,g^{[^3S_1]}\sqrt{\mathcal{Z}}\,\frac{C_0(\eta_p)W_1^{1/2}(\eta_p)}{K^{[\xi]}(p)-H_1(\eta_p)}\,e^{ i\sigma_1}\,\varepsilon^d_i\,\varepsilon^{Li*}_{j}\,(\boldsymbol{\varepsilon}^{\gamma*}\cdot\hat{\mathbf{p}}).
\end{eqnarray}
with
%\begin{eqnarray}\label{eq:41}
%\nabla_{\!k}\big[P_1(\mathbf{\hat{p}}\cdot\mathbf{\hat{r}})\chi^{(1)}_p(r)\big]=ie^{ i\sigma_1(\eta_{p})}\hat{\mathbf{p}}_l \Big(\delta_{lk}+\frac{r_lr_k}{r}\frac{\partial}{\partial r}\Big)\frac{F_1(\eta_p ,pr)}{pr^2},
%\end{eqnarray}
\begin{eqnarray}\label{eq:43}
\,\,\,\mathcal{A}_1(p)=\frac{\Gamma (1\!+\!k_C/\gamma)}{C_0(\eta_p)}\!\int_0^\infty \!\! d r r W_{-k_C/\gamma,1/2}(2\gamma r)\Big(3+r\frac{\partial}{\partial r}\Big)\frac{F_1(\eta_p ,pr)}{pr^2},
\end{eqnarray}
%\begin{eqnarray}\label{eq:45}
%\lim_{\mathbf{r}''\rightarrow 0}\mathbf{\nabla}_{\!k}\mathbf{\nabla}^{''}_{\!j}[P_1(\hat{\mathbf{r}}\cdot \hat{\mathbf{r}}'')G_C^{(1)}(E,r,r'')]
%&\!\!\!=&\!\!\!\frac{i\mu  p  }{6 \pi}\Gamma(2\!+\!i\eta_p)
%\nonumber\\
%\!\!\!\!\!\!\!\!\!\!\!\!\!\!\!\!\!\!\!\!\!\!\!\!\!\!&&\!\!\!\!\!\!\!\!\!\!\!\!\!\!\!\!\!\!\!\!\!\!\!\!\!\!\!\!\!\!\!\!\!\!\!\!\!\times\Bigg[\delta_{jk}\frac{W_{-i\eta _p,3/2}(-2ipr)}{r^2}+\frac{r_kr_j}{r} \frac{\partial }{\partial r}\frac{W_{-i\eta _p,3/2}(-2ipr)}{r^2}\Bigg]\quad\,\,\,\,.
%\end{eqnarray}
\begin{eqnarray}\label{eq:44}
\mathcal{B}_1(p)=ip\,\Gamma (1\!+\!k_C/\gamma)\,\Gamma(2\!+\!i\eta_p)\!\int_0^{\infty}\!\!dr r  W_{-k_C/\gamma , 1/2}(2\gamma r)\Big(\frac{2}{r} +\frac{\partial }{\partial r} \Big)\frac{W_{-i\eta _p,3/2}(-2ipr)}{r}.
\end{eqnarray}
We are working in Coulomb gauge where the relation $\varepsilon^{\gamma}\cdot \mathbf{k}=0$ for the real photon with momentum $\mathbf{k}$ is fulfilled. Also, the spherical symmetry $r_ir_j/r^2\!\rightarrow \!\delta_{ij}/3$ is used in the integrals of Eqs.~\ref{eq:39} and~ \ref{eq:40}.
The Whittaker function $W_{-k_C/\gamma,1/2}(2\gamma r)$, in diagrams $a_1$, $a_2$, $b_1$ and $b_2$ is associated with the final $S$-wave bound state wave function and the $P$-wave Coulomb wave function $F_1(\eta_p ,pr)$ in $a_1$ and $a_2$ diagrams corresponds to the initial incoming scattering state \cite{abromowitz1972handbook}. Moreover, due to the presence of one
bound-state field, there is a wave function renormalization $\sqrt{\mathcal{Z}}$ present.
Thus, the $E1$ transition amplitude for the $d(\alpha, \gamma){^6\textrm{Li}}$ process is summarized as
\begin{eqnarray}\label{eq:45}
\mathcal{M}_{E1}(p)&\!\!\!\!=&\!\!\!\!2\,Q_{1}\,g^{[^3S_1]}\sqrt{\mathcal{Z}}C_0(\eta_p)\,e^{ i\sigma_1}\!\Bigg[\!\mathcal{A}_1(p)\!+\! \frac{W_1^{1/2}(\eta_p)}{K^{[\xi]}(p)-H_1(\eta_p)}\Big(\mathcal{B}_1(p)\!-\!\frac{3\pi i}{\mu}J_0  (i\gamma)\!-\!3\pi i k_0L_{E1}\!\Big)\!\Bigg].\nonumber\\
\end{eqnarray}

By increasing the CM energy, the $E2$ transition would be more important in the $d(\alpha , \gamma )^6{\textrm{Li}} $ reaction. So, we should consider the contribution of the $E2$ transition to include $3^+$ resonance at $E_{CM}=0.711$ MeV.
So, in the following, the $E2$ transition for $d(\alpha,\gamma)^6{\textrm{Li}}$ reaction which involves transition from $^3D_1$, $ ^3D_2$  and $^3D_3$ to $^3S_1$ would be considered. The contribution of all diagrams in the $E2$ transition can be summarized as
\begin{eqnarray}\label{eq:46}
M_{E2}(\mathbf{p})=\!\!\!\!\!\!\!\!\!\sum_{\xi={^3\!D_1}, {^3\!D_2}, {^3\!D_3}}\!\!\!\!\!\!\!\big(M_{E2,a_{1,2}}^{[\xi]}(\mathbf{p})\!+M_{E2,b_{1,2}}^{[\xi]}(\mathbf{p})\!+M_{E2,b_3}^{[\xi]}(\mathbf{p})\!+M_{E2,c}^{[\xi]}(\mathbf{p})\big),
\end{eqnarray}
with
\begin{eqnarray}\label{eq:47}
M_{E2}(\mathbf{p})=\mathcal{M}_{E2}(p)\,\varepsilon^d_i \,\varepsilon^{Li*}_{j} (\mathbf{\hat{k}}\cdot\hat{\mathbf{p}})(\boldsymbol{\varepsilon}^{\gamma*}\cdot\hat{\mathbf{p}}).
\end{eqnarray}
The $E2$ transition amplitude from the initial ${^3\!D_1},{ ^3\!D_2},$ and $ {^3\!D_3}$ states to the final bound state $^3\!S_1$, for the diagrams $a_1$, $a_2$, $b_1$, $b_2$ and $b_3$ are obtained as
\begin{eqnarray}\label{eq:48}
M_{E2,a_{1,2}}^{[\xi]}(\mathbf{p})&\!\!\!\!\!\!=&\!\!\!\!-\frac{1}{\mu}Q_{2}\,g^{[^3S_1]}\sqrt{\mathcal{Z}}\,   \,\varepsilon^d_i\,\varepsilon^{Li*}_{j}\,\varepsilon^{\gamma*}_k  k_l \, \!\int\! d^3 r\, G_C^{(0)}(-B,0,r) r_l\nabla_{\!k}\big[5P_2(\mathbf{\hat{p}}\cdot\mathbf{\hat{r}})\chi^{(2)}_p(r)\big],\nonumber\\&\!\!\!\!\!\!=&\!\!\!\!-2 k\,Q_{2}\,g^{[^3S_1]}\sqrt{\mathcal{Z}}\,e^{ i\sigma_2} \mathcal{A}_2(p)\,\varepsilon^d_i \,\varepsilon^{Li*}_{j} (\mathbf{\hat{k}}\cdot\hat{\mathbf{p}})(\boldsymbol{\varepsilon}^{\gamma*}\cdot\hat{\mathbf{p}}),\\\nonumber\\
M_{E2,b_{1,2}}^{[\xi]}(\mathbf{p})&\!\!\!\!\!=&\!\!\! \frac{i}{\mu}Q_{2}\,g^{[^3S_1]}\sqrt{\mathcal{Z}}\,   \,\varepsilon^d_i\,\varepsilon^{Li*}_{j}\,\varepsilon^{\gamma*}_k\,\frac{5T^{[\xi]}_{CS}(p)e^{i\sigma_2}}{C_0(\eta_p)W_2^{1/2}(\eta_p)}(\hat{\mathbf{p}}_n\hat{\mathbf{p}}_m-\frac{1}{3}\delta_{nm})k_l\int \!\!d^3 r\,\nonumber\\
&&\!\!\!\times G_C^{(0)}(-B,0,r) r_l \nabla_{\!k}\!\lim_{\mathbf{r}''\rightarrow 0}(\nabla^{''}_n\nabla^{''}_m\!-\!\frac{1}{3}\delta_{nm}\nabla^{''}_n\nabla^{''}_m)[5P_2(\hat{\mathbf{r}}\cdot \hat{\mathbf{r}}'')G_C^{(2)}(E,r,r'')]\nonumber\\
&\!\!\!\!\!=&\!\!\! -\frac{2i}{3}k Q_{2}\,g^{[^3S_1]}\sqrt{\mathcal{Z}}\,\frac{C_0(\eta_p)W_2^{1/2}(\eta_p)}{K^{[\xi]}(p)-H_2(\eta_p)}\,e^{ i\sigma_2} \mathcal{B}_2(p)\,\varepsilon^d_i \,\varepsilon^{Li*}_{j} (\mathbf{\hat{k}}\cdot\hat{\mathbf{p}})(\boldsymbol{\varepsilon}^{\gamma*}\cdot\hat{\mathbf{p}}),\\\nonumber\\
M_{E2,b_3}^{[\xi]}(\mathbf{p})&\!\!\!=&\!\!\!-Q_{2}\,g^{[^3S_1]}\sqrt{\mathcal{Z}}\,\varepsilon^d_i \,   \varepsilon^{Li*}_{j} \,\varepsilon^{\gamma*}_k\,k_l \,G_C^{(0)}(-B,0,0)\, \frac{5T^{[\xi]}_{CS}(p)e^{i\sigma_2}}{C_0(\eta_p)W_2^{1/2}(\eta_p)} (\hat{\mathbf{p}}_k\hat{\mathbf{p}}_l-\frac{1}{3}\delta_{kl}) \nonumber\\&\!\!\!=&\!\!\!\frac{10\pi }{\mu} k\,Q_{2}\,g^{[^3S_1]}\sqrt{\mathcal{Z}}\,\frac{C_0(\eta_p)W_2^{1/2}(\eta_p)}{K^{[\xi]}(p)-H_2(\eta_p)}\,e^{ i\sigma_2}J_0  (i\gamma)\,\varepsilon^d_i \,\varepsilon^{Li*}_{j} (\mathbf{\hat{k}}\cdot\hat{\mathbf{p}})(\boldsymbol{\varepsilon}^{\gamma*}\cdot\hat{\mathbf{p}}),\\\nonumber\\
M_{E2,c}^{[\xi]}(\mathbf{p})&\!\!\!=&\!\!\!-\mu\,Q_{2}\,g^{[^3S_1]}\sqrt{\mathcal{Z}}\, \,\varepsilon^d_i  \,\varepsilon^{Li*}_{j} L_{E1}\,k_0 \, \varepsilon^{\gamma*}_kk_l\frac{5T^{[\xi]}_{CS}(p)e^{i\sigma_2}}{C_0(\eta_p)W_2^{1/2}(\eta_p)}(\hat{\mathbf{p}}_k\hat{\mathbf{p}}_l-\frac{1}{3}\delta_{kl})\nonumber\\&\!\!\!=&\!\!\!
10\pi\,k k_0\,Q_{2}\,g^{[^3S_1]}\sqrt{\mathcal{Z}} L_{E2}\frac{C_0(\eta_p)W_2^{1/2}(\eta_p)}{K^{[\xi]}(p)-H_2(\eta_p)}\,e^{ i\sigma_2}\,\varepsilon^d_i \,\varepsilon^{Li*}_{j} (\mathbf{\hat{k}}\cdot\hat{\mathbf{p}})(\boldsymbol{\varepsilon}^{\gamma*}\cdot\hat{\mathbf{p}}),
\end{eqnarray}
with
%\begin{eqnarray}\label{eq:50}
%\nabla_{\!k}\big[P_2(\mathbf{\hat{p}}\cdot\mathbf{\hat{r}})\chi^{(2)}_p(r)\big]=-e^{i\sigma_2(\eta_p)}\Bigg [ \frac{3}{r}(\hat{\mathbf{p}}\cdot \hat{\mathbf{r}})\bigg(\mathbf{\hat{p}}_k-\hat{\mathbf{r}}_k(\hat{\mathbf{p}}\cdot \hat{\mathbf{r}})\bigg)\!+\!P_2(\hat{\mathbf{p}}\cdot \hat{\mathbf{r}})\hat{\mathbf{r}}_k\frac{\partial}{\partial r}\Bigg]\frac{F_2(\eta_p,pr)}{pr},
%\end{eqnarray}
\begin{eqnarray}\label{eq:52}
\mathcal{A}_2(p)= \frac{\Gamma (1\!+\!\frac{k_C}{\gamma})}{C_0(\eta_\eta)}\,\!\int_0^\infty\!\! \!d rr^2\, W_{-k_C/\gamma,1/2}(2\gamma r)\Big ( \frac{3}{r}\!+\!\frac{\partial}{\partial r}\Big)\frac{F_2(\eta_p,pr)}{pr},\,\,\,\,\,
\end{eqnarray}
%\begin{eqnarray}\label{eq:53}
%\lim_{\mathbf{r}''\rightarrow 0}\mathbf{\nabla}_k(\nabla^{''}_n\nabla^{''}_l-\frac{1}{3}\delta_{nl}\nabla^{''}_m\nabla^{''}_m)[P_2(\hat{\mathbf{r}}\cdot \hat{\mathbf{r}}'')G_C^{(2)}(E,r,r'')]=\frac{\mu  p^2  }{60 \pi}\Gamma(3+i\eta_p)\quad\quad\quad\quad\quad\quad\nonumber\\
%\times \bigg[3(\delta_{kn}\hat{\mathbf{r}}_l\!+\!\delta_{lk}\hat{\mathbf{r}}_n\!-\!2\,\hat{\mathbf{r}}_n\hat{\mathbf{r}}_l\hat{\mathbf{r}}_k)\frac{W_{-i\eta _p,5/2}(-2ipr)}{r^2}\!+\!(3\hat{\mathbf{r}}_n\hat{\mathbf{r}}_l\hat{\mathbf{r}}_k\!-\!\delta_{nl}\hat{\mathbf{r}}_k)\frac{\partial}{\partial r}\frac{W_{-i\eta _p,5/2}(-2ipr)}{r}\bigg].
%\end{eqnarray}
\begin{eqnarray}\label{eq:53}
\mathcal{B}_2(p)= -p^2\Gamma (1\!+\!\frac{kc}{\gamma})\,\Gamma(3\!+\!i\eta_p)\!\int_0^\infty\!\! drr^2\,W_{-k_C/\gamma,1/2}(2\gamma r)\Big(\frac{3}{r}+\frac{\partial}{\partial r}\Big)\frac{W_{-i\eta _p,5/2}(-2ipr)}{r}.
\end{eqnarray}

Spherical symmetry is used to write $ r_kr_jr_ir_l/r^4 \!\rightarrow \!(\delta_{kj}\delta_{il}+\delta_{ki}\delta_{jl}+\delta_{kl}\delta_{ji})/15$ in the integrals.
Thus, the $E2$ transition amplitude for the $d(\alpha, \gamma){^6\textrm{Li}}$ capture reaction from the initial $D$-wave states is summarized as
\begin{eqnarray}\label{eq:54}
\mathcal{M}^{[\xi]}_{E2}(p)&\!\!\!\!\!\!=&\!\!\!\!-2k\,Q_{2}\,g^{[^3S_1]}\sqrt{\mathcal{Z}}C_0(\eta_p)\,e^{ i\sigma_2}\nonumber\\&&\!\!\!\!\times\Big[\mathcal{A}_2(p)\!+\! \frac{W_2^{1/2}(\eta_p)}{K^{[\xi]}(p)-H_2(\eta_p)}\Big(\frac{i}{3 } \,\mathcal{B}_2(p)\!-\!\frac{5\pi}{\mu}J_0  (i\gamma)\!-\!5\pi  k_0L_{E2}\Big)\!\Big].
\end{eqnarray}
The loops of the diagrams
$b_1$ and $b_2$ lead to a logarithmic divergence
in $M_{E1,b_{1,2}}^{[\xi]}$ and $M_{E2,b_{1,2}}^{[\xi]}$ amplitudes when $r$ goes to zero.
A short-range cutoff $r_C$ in the $r$ integral in Eqs. \ref{eq:44} and \ref{eq:53} is introduced as a regulator and the divergence can be handled by renormalizing the counter terms $L_{E1}$ and $L_{E2}$ \cite{ando2021cluster}, \cite{ando2019s}. The loop of the
diagram $b_3$ is also divergent and could be renormalized by the $L_{E1}$ and $L_{E2}$ terms as well. By extracting the divergent parts, the renormalized $L_{E1}$ and $L_{E2}$ can be expressed as \begin{eqnarray}\label{eq:55}
L_{E1}^R=L_{E1}+\frac{1}{\mu k_0}\bigg[J_0^{div}+\frac{i \mu}{3\pi}\mathcal{B}_1^{div}\bigg],
\end{eqnarray}
\begin{eqnarray}\label{eq:56}
L_{E2}^R=L_{E2}+\frac{1}{\mu k_0}\bigg[J_0^{div}-\frac{i\mu }{15\pi}\mathcal{B}_2^{div}\bigg],
\end{eqnarray}
%\begin{eqnarray}\label{eq:57}
%L_{E1}^R=L_{E1}+\bigg[J_0^{div}-\frac{i\mu p}{2\pi}\mathcal{B}_2^{div}\bigg],
%\end{eqnarray}
with
\begin{eqnarray}\label{eq:57}
\mathcal{B}_1^{div}=\mathcal{B}_2^{div}=k_C\int_0^{r_C}\!\frac{dr}{r}\rightarrow k_C(\frac{\kappa}{2})^{2\epsilon}\int_0^{r_C}\frac{dr}{r^{1-2\epsilon}}=k_C\bigg[\frac{1}{2\epsilon}+\ln(\frac{\kappa }{2}r_C)+\mathcal{O}(\epsilon)\bigg]. %r^{-1+2\epsilon}=\frac{1}{2\epsilon}+\ln(\mu_{DR}}{2}r_C)+\mathcal{O}(\epsilon),
\end{eqnarray}
In the power divergence subtraction (PDS) regularization scheme, the divergent part of $J_0$ that does not depend on the momentum is obtained as \cite{kong2001proton}
\begin{eqnarray}\label{eq:58}
J_0^{div}=\frac{\mu k_C}{2\pi}\bigg[\frac{1}{\epsilon}+\textrm{ln}\bigg(\frac{\pi\kappa^{2}}{4k^{2}_C}\bigg)\!+2-3\,C_E\bigg],
\end{eqnarray}
with $D$ the dimensionality of spacetime, $\kappa$ the renormalization
mass scale and $C_E=0.577...$ Euler–Mascheroni constant.
The $L^R_{E1}$ and $L^R_{E2}$ refer to the renormalized parameters that have been adjusted with fitting to the astrophysical S-factor experimental data.

\section{Results}\label{Results}
The astrophysical S-factor plays a crucial role in the nuclear fusion reactions in the core of stars. This factor allows us to calculate how quickly those reactions take place, taking into account the interaction between the reacting particles. It defines as
\begin{equation}\label{eq:62}
S(E) \equiv E \; \textrm{exp}(2 \pi \eta_p) \; \sigma(p),
\end{equation}
where $E=p^2/2\mu$ is the CM energy of system and $\sigma(p)$ denotes the total cross-section. The differential cross section for our reaction is calculated by averaging over the initial spin
and summing over all components of the final state i.e., the outgoing photon polarization and the $^6\mathrm{Li}$ spin components as
\begin{eqnarray}\label{eq:63}
\frac{d\sigma}{d\Omega}=\frac{\mu k}{8\pi^2p}\frac{1}{9}\sum^3_{i,j=1}\sum^2_{r=1}\Big|\big[\mathcal{M}_{E1}(p)+\mathcal{M}_{E2}(p) (\mathbf{\hat{k}}\cdot\hat{\mathbf{p}})\big](\boldsymbol{\varepsilon}_r^{\gamma*}\cdot\hat{\mathbf{p}})\,\varepsilon^d_i \,\varepsilon^{Li*}_{j} \Big|^2,
\end{eqnarray}
by considering photon momentum direction $\mathbf{\hat{k}}$ along $\hat{z}$ axis and integrating over angle variables, the total cross section is presented by
\begin{eqnarray}\label{eq:63}
\sigma(p)=\frac{\mu k}{18\pi p}\Big(|\mathcal{M}_{E1}(p)|^2+\frac{1}{5}|\mathcal{M}_{E2}(p)|^2\Big).
\end{eqnarray}
To calculate the S-factor using the EFT expressions for the $E1$ and $E2$ transition amplitudes of Eqs.~\ref{eq:45} and \ref{eq:54}, we need to determine the values of the elastic $P$- and $D$- waves scattering parameters and the $E1$ and $E2$ two-body coupling constants. We recently obtained $S$-, $P$-, and $D$-wave scattering parameters  by phase shift analysis in our previous paper~\cite{Nazari2023low}.
For the $E1$ and $E2$ two-body coupling constants $L^R_{E1}$ and $L^R_{E2}$, we have matched our EFT relation for the S-factor to the available experimental
data~\cite{trezzi2017big,igamov1999triple,kiener1991measurements,mohr1994direct,anders2014first,robertson1981observation} at the energy range $0.001-3.3 \, \textrm{MeV}$ with arbitrary values of the cutoff $r_C$.
Table~\ref {tab:table2} shows the results of the S-factor in energies $E_{CM} = \,$0.2 and 3\,\,$\textrm{MeV}$ as a function of $r_C$. The $L^R_{E1}$ and $L^R_{E2}$ constants and S-factor demonstrate a notable correlation with the short-range cutoff $r_C$. As the radius $r_C$ decreases, the S-factor increases gradually. It can be seen that for the cutoff values below $r_C=0.1$ fm, the S-factor values would be stable. This indicates the insensitivity of the S-factor to the precisely chosen short-range cutoff within the region of $r_C\leq 0.1$ fm.
\begin{table}\centering
\caption{\label {tab:table2}Fitted values of the $L^R_{E1}$ and $L^R_{E2}$ constants to experimental data of S-factor
at $E_{\textrm{CM}} = 0.001-3.3 \,\textrm{MeV}$ with the cutoff $r_C = 0.001-1 \,\textrm{fm}$. In the fourth and the last columns, the results
of S-factor at $E_{CM} = 0.2  \,\textrm{MeV}$ and $E_{CM} = 3  \,\textrm{MeV}$ are presented.}
%\begin{ruledtabular}
\begin{tabular}{|c|c|c|c|c|}
\hline
 $ r_C$(fm)&$L^R_{E1}$& $L^R_{E2}$&S-factor(MeV.b)&S-factor(MeV.b)\\
&&&$(E_{CM}=0.2\,\,\textrm{MeV})$&$(E_{CM}=3\,\,\textrm{MeV})$\\
\hline
\hline
1&$(5.805\pm 0.881)\times 10^{-1}$&$2.197 \pm 0.383$&$4.841\times 10^
{-9}$&$3.271 \times 10^{-7}$\\
\hline
0.5&$(5.421\pm 0.732)\times 10^{-1}$&$1.945 \pm 0.326$&$4.985\times 10^
{-9}$&$3.724 \times 10^{-7}$\\
\hline
0.2&$(5.187\pm 0.865)\times 10^{-1}$&$1.621\pm 0.293$&$5.171\times 10^
{-9}$&$3.993 \times 10^{-7}$\\
\hline
0.1&$(4.942\pm 0.719)\times 10^{-1}$&$1.352 \pm 0.318$&$5.312\times 10^
{-9}$&$4.083 \times 10^{-7}$\\
\hline
0.05&$(4.733\pm 0.748)\times 10^{-1}$&$1.349\pm 0.304$&$5.319\times 10^
{-9}$&$4.101 \times 10^{-7}$\\
\hline
0.001&$(4.721\pm 0.752)\times 10^{-1}$&$1.344\pm 0.315$&$5.320\times 10^
{-9}$&$4.103 \times 10^{-7}$\\
\hline
\end{tabular}
%\end{ruledtabular}
\end{table}

According to the power counting proposed in Sec.~\ref{E1 and E2 Transition Amplitudes}, the dominant contributions of the scattering amplitude in channels $ ^3P_0$, $ ^3P_1$, $^3P_2$, $^3D_1$ and $^3D_2$ come clearly from their scattering lengths and the influences of both their effective ranges and shape parameters are small and can be considered as higher-order corrections~\cite{Nazari2023low}. Based on this power counting, the effective charges $Q_1$ and $Q_2$ scale as $Q^4/\Lambda^4$ and $Q/\Lambda$, respectively. We also consider $\mathcal{B}_2/\mathcal{A}_2\sim \Lambda^4/Q^2$ and $\mathcal{B}_1/\mathcal{A}_1\sim Q^2 \Lambda$ which hold over a range of $0.1\lesssim E_{CM}\lesssim 3.3$ MeV. Taking into account these analyses, we can estimate the order of all diagrams in both $E1$ and $E2$ transition amplitudes for each channel. In Table~\ref{tab:3}, the relative contribution of diagrams $b_{1,2}$, $b_3$ and $c$ for each channel to the diagrams $a_{1,2}$ in channel $^3D_3$ are presented.
\begin{table}[h]
\centering
\caption{\label{tab:3} The relative contribution of diagrams $b_{1,2}$, $b_3$, $c$ for each channel to the diagrams $a_{1,2}$ for channel $^3D_3$ around the resonance energy $E_{CM}=0.71$ MeV, according to the suggested power counting.   }
\begin{tabular}{|c|c|c|c|c|c|c|c|}
\hline
$\xi$ &$\mathcal{M}_{E_1,a_{1,2}}^{[\xi]}/\mathcal{M}_{E_2,a_{1,2}}^{[\xi]}$ &$\mathcal{M}_{E_1,b_{1,2}}^{[\xi]}/\mathcal{M}_{E_2,a_{1,2}}^{[\xi]}$ &$\mathcal{M}_{E_1,b_3}^{[\xi]}/\mathcal{M}_{E_2,a_{1,2}}^{[\xi]}$& $\mathcal{M}_{E_1,c}^{[\xi]}/\mathcal{M}_{E_2,a_{1,2}}^{[\xi]}$\\
\hline
\hline
$^3P_0$&  $Q/\Lambda$&  $Q/\Lambda$&$Q/\Lambda$  &$Q/\Lambda$ \\
\hline
$^3P_1$& $Q/\Lambda$& $Q^2/\Lambda^2$&$Q^2/\Lambda^2 $&$Q^2/\Lambda^2$ \\
\hline
$^3P_2$&  $Q/\Lambda$& $Q^2/\Lambda^2$&$Q^2/\Lambda^2 $&$Q^2/\Lambda^2$ \\
\hline
\hline
$\xi$ &$\mathcal{M}_{E_2,a_{1,2}}^{[\xi]}/\mathcal{M}_{E_2,a_{1,2}}^{[\xi]}$ &$\mathcal{M}_{E_2,b_{1,2}}^{[\xi]}/\mathcal{M}_{E_2,a_{1,2}}^{[\xi]}$ &$\mathcal{M}_{E_2,b_3}^{[\xi]}/\mathcal{M}_{E_2,a_{1,2}}^{[\xi]}$& $\mathcal{M}_{E_2,c}^{[\xi]}/\mathcal{M}_{E_2,a_{1,2}}^{[\xi]}$\\
\hline
\hline
$^3D_1$&  1& $Q^4/\Lambda^4$ &$Q/\Lambda $&$Q/\Lambda $  \\
\hline
$^3D_2$&  1& $Q^4/\Lambda^4$ &$Q/\Lambda $&$Q/\Lambda $ \\
\hline
$^3D_3$&  1& $Q^2/\Lambda^2$ &$\Lambda/Q $&$\Lambda/Q $  \\
\hline
\end{tabular}
\end{table}
It is shown the leading-order contribution comes from the diagrams $b_3$ and $c$ in the $^3D_3$ channel that can independently reproduce the resonance of the S-factor observed at $E_{CM}= 0.711 \,\,\textrm{MeV}$. Around the resonance energy, the $a_{1,2}$ diagrams for all $D$-waves can be considered as the next-to-leading-order corrections. Additionally, the $b_3$ and $c$ diagrams corresponding to the $^3D_1$ and $^3D_2$ waves along with the $a_{1,2}$ diagrams for all $P$ waves have $\mathrm{N^{2}LO}$ contribution same as the $b_{1,2}$, $b_3$ and $c$ diagrams for the $^3P_0$ channel. The remaining diagrams in Table~\ref{tab:3} are affected by S-factor as higher-order corrections.

\begin{figure}
\begin{center}
\includegraphics[width=5.2in,height=8.6cm]{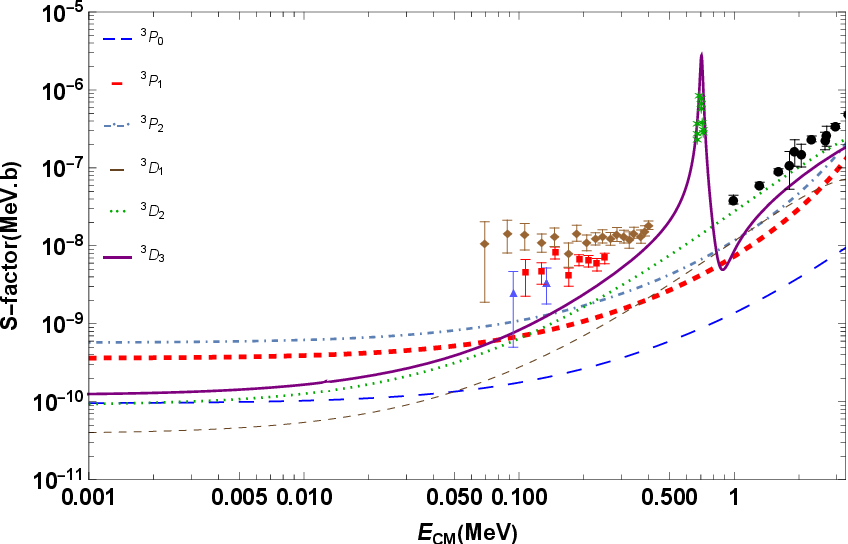}
\end{center}
\caption{Contributions of the different incoming partial waves components in our calculated EFT astrophysical
S-factor for the capture reaction $d(\alpha,\gamma)^6{\textrm{Li}}$ compared with the experimental data of direct measurements: black circles \cite{robertson1981observation}, blue triangles\cite {anders2014first}, green stars \cite{mohr1994direct}, brown diamonds \cite{kiener1991measurements}, red squars\cite{igamov1999triple}.}\label{fig:4}
\end{figure}

\begin{figure}
\begin{center}
\includegraphics[width=5.2in,height=8.6cm]{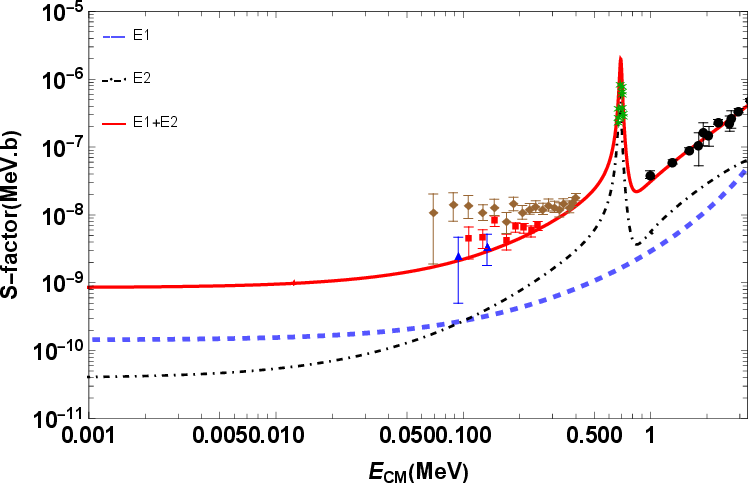}
\end{center}
\caption{The contribution of the $E1$ and $E2$, and the $E1$ + $E2$ transitions to the astrophysical S-factor for the capture reaction $d(\alpha,\gamma)^6{\textrm{Li}}$ according to our  cluster effective field theory approach. In this figure, the order of each diagram is also shown. The experimental data are the same as Fig. \ref{fig:4}. }\label{fig:5}
\end{figure}

In Fig. \ref{fig:4}, the contributions of each partial wave in our EFT calculation for the astrophysical S-factor of the capture reaction $d(\alpha,\gamma)^6{\textrm{Li}}$ are presented.
It shows that the primary contribution comes from the initial $^3P_2$ partial wave at energies below $0.1 \,\,\textrm{MeV}$ in the CM framework. Moving into the resonance region, the $^3D_3$ channel becomes the dominant contribution making the resonance at $E_{CM}=0.711\,\,\textrm{MeV}$. The EFT results indicated in Fig. \ref{fig:4} are in agreement with the power counting estimations in Table~\ref{tab:3}. Also, the contributions of $E1$, $E2$, and the $E1 + E2$ transitions of the astrophysical S-factor for the capture
reaction $d(\alpha,\gamma)^6{\textrm{Li}}$ are shown in Fig.~\ref{fig:5}.

In the two-body models, the $E1$ transition from $P$-waves to the $^6\text{Li}$ ground state is significantly hindered by the isospin selection rule for $N \!=\! Z$ nuclei due to the nearly identical charge-to-mass ratios of the deuteron and the $\alpha$ particle.

This appears as a large suppression factor $(Z_d /m_d -Z_{\alpha}/m_{\alpha})$ in the $E1$ transition amplitude. According to the incoming alpha particle $( J^{\pi}= 0^+$ and 
$I = 0 )$ and the deuteron $( J^{\pi}= 1^+$ and $I = 0 )$ and also the final $^6{\textrm{Li}}$ nucleus in its ground state$( J^{\pi}= 1^+$ and $I = 0 )$, and considering the point-like treatments for the alpha and deuteron, the dominant iso-scalar part of the $E1$ transition only contributes in the calculation of the $E1$ transition amplitude. The iso-vector part comes from excited states of incoming particles or the final state, which can be considered at higher energy. Since there is no state of substructure for the deuteron as $n + p$ for the triplet isospin contribution, the $d-\alpha$ system’s isospin is zero. Consequently, there is no isospin breaking in the current CEFT. Here, the effects of the isospin mixture can be considered in $d(\alpha,\gamma)^6{\textrm{Li}}$ by inclusion of the $n -^5{\textrm{Li}}$ and $p - ^5{\textrm{He}}$ systems in the intermediate states at high energy. 

At energies 4.19 $\textrm{MeV}$ and 3.18 $\textrm{MeV}$ above the $d- \alpha$ breakup threshold, the $n -{^5{\textrm{Li}}}$ and $p - {^5{\textrm{He}}}$ breakup channels respectively, are open. In this case, we might have introduced the $n -{^5{\textrm{Li}}}$ and $p - {^5{\textrm{He}}}$ fields as relevant degrees of freedom in the theory. The $n -{^5{\textrm{Li}}}$ and $p - {^5{\textrm{He}}}$ fields then appear in the intermediate states as $n -{^5{\textrm{Li}}}$ and $p - {^5{\textrm{He}}}$ propagation in the loop diagrams $b_1$, $b_2$, and $b_3$ instead of the $d-\alpha$ propagation. 
In our work, however, the $n -{^5{\textrm{Li}}}$ and $p - {^5{\textrm{He}}}$ fields are regarded as irrelevant degrees of fr eedom at high energy and are integrated out of the effective Lagrangian. This effect is embedded in the coefficient and the contact interaction $L^R_{E1}$ in diagram $c$, which are fitted to the experimental data of the S-factor.

\begin{figure}
\begin{center}
\includegraphics[width=5.2in,height=8.6cm]{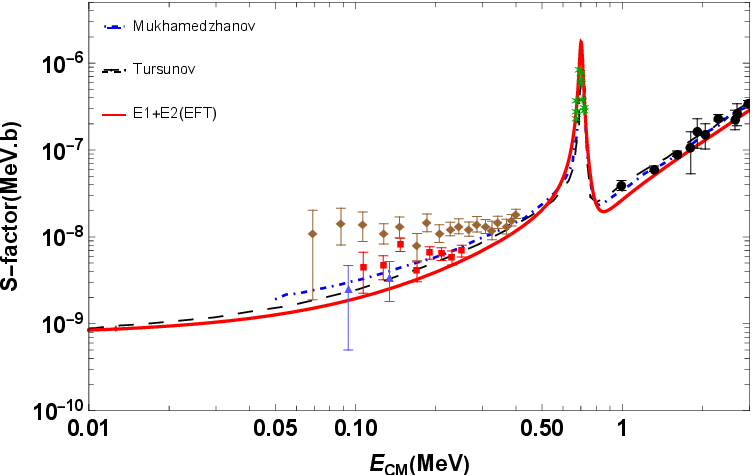}
\end{center}
\caption{Comparison of the result obtained for astrophysical S-factor from cluster EFT approach with the results obtained from two other theoretical approaches~\cite{mukhamedzhanov2011reexamination,tursunov2018theoretical}.}
\label{fig:6}
\end{figure}

As we expect, the $E1$ capture will be more dominant than $E2$ capture at energies lower than $0.1 \,\textrm{MeV}$ due to the different energy dependencies of the penetration probabilities in the $P$- and $D$-waves.
In contrast, the $E2$ component becomes more significant at energies related to the resonance energy and higher. At energies below 0.1 MeV, it is expected that $E1$ capture will be more dominant than $E2$ capture in the center-of-mass frame due to the distinct energy dependencies of the penetration probabilities in $P$- and $D$- waves \cite{hammache2010high}.
Finally, the result obtained for astrophysical S-factor based on cluster EFT approach has been compared with the results obtained from two other theoretical approaches in Fig.~\ref{fig:6} ~\cite{mukhamedzhanov2011reexamination,tursunov2018theoretical}.

\section{Conclusion}\label{Conclusion}
In this paper, we have studied the low-energy $d(\alpha,\gamma)^6{\textrm{Li}}$ gamma capture reaction with the two-body cluster EFT approach. Based on this approach, the deuteron and alpha nucleus are
the point-like nuclear clusters and considering the fact that there are no existing explicit pions, we have concentrated on
the energy region $E_{CM}\lesssim 3.3 \,\,\textrm{MeV}$. In this energy region, the Coulomb force was considered nonperturbatively. 
In the study of the radiative  $d(\alpha,\gamma)^6{\textrm{Li}}$ capture reaction, we calculate
the radiative capture amplitudes for
the initial $P$- and $D$-waves of the $d-\alpha$ system.The Isoscaler component of $E1$ transition from initial $P$-wave states to the ground state of $^6{\textrm{Li}}$ was considered and the $E2$ transition contains $D$-wave initial states.

Our EFT results in Fig. \ref{fig:4} illustrate the significance of $l \leq 2 $ partial waves in the astrophysical S-factor of the reaction where a deuterium nucleus captures an alpha particle to form $^6\textrm{Li}$. 
It is evident that at energies below 0.1 MeV in the CM framework, the primary contribution comes from the initial $^3P_2$ partial
wave. Moving into the resonance region, the $^3D_3$ channel becomes the dominant contribution,
making the resonance at $E_{CM}= 0.711\,\, \textrm{MeV}$.
Below an energy threshold of $0.1\,\, \textrm{MeV}$ in the CM frame, the primary influence stems from the initial $^3P_2$ partial wave. 

Next, we analyzed the cluster Effective Field Theory results for the $E1$ and $E2$ contributions, contrasting them with existing experimental data. In Fig. \ref{fig:5}, the $E1$, $E2$, and combined $E1$+$E2$ transitions of the astrophysical S-factor for the capture reaction $d(\alpha,\gamma)^6\text{Li}$ are represented. 
The primary role of the $E1$ transition is emphasized in the astrophysical S-factor of the gamma capture reaction $d(\alpha,\gamma)^6{\textrm{Li}}$ for energies below $0.1 \,\textrm{MeV}$. Once the energy surpasses this threshold, the dominance shifts towards the $E2$ channel. It has been determined that the $E2$ transition, specifically through the two-body current diagram, plays a significant role in accurately capturing the resonance behavior observed at energies above $0.1 \,\textrm{MeV}$. 

In the current CEFT calculation, where the deuteron is assumed to be a point-like particle, the results for $E_{CM} > 3.3 \,\textrm{MeV}$ are controversial. Employing the three-body cluster formalism is crucial for accurately understanding and calculating cross sections at higher energies. To resolve the disparities in the S-factor findings for CM energies exceeding $3.3 \,\textrm{MeV}$, a three-body cluster effective field theory method can be utilized. This approach considers the neutron, proton, and alpha particles as relevant degrees of freedom.

\section*{Acknowledgment}
This work is based upon research funded by the Iran National Science Foundation (INSF) under project No. 4003662.
We are grateful to Dr. Ergash Tursunov for sharing the experimental data.

\newpage
% BibTeX users please use
% \bibliographystyle{}
% \bibliography{}
%
% Non-BibTeX users please use
%\bibliographystyle{ptephy}
\bibliography{Ref7}

\end{document}